\documentclass[aps,prl,twocolumn,floatfix,superscriptaddress]{revtex4}
\usepackage[]{graphicx}
\usepackage{epstopdf}

\usepackage{graphics}
\usepackage{subfigure}
\usepackage{amsmath}
\usepackage{physics}
\usepackage{amsfonts}
\usepackage{amssymb}
\usepackage{float}
\usepackage{longtable}
\usepackage{epsfig}
\usepackage{latexsym}
\usepackage{theorem}
\usepackage{bbm}

\usepackage{dcolumn}
\usepackage{bm}
\usepackage[usenames]{color}    

\usepackage{siunitx}
\usepackage{enumerate}

\begin{document}

\title{Experimental quantum reading with photon counting}

\author{Giuseppe Ortolano}
\email{giuseppe.ortolano@polito.it}
\affiliation{Quantum metrology and nano technologies division,  INRiM,  Strada delle Cacce 91, 10153 Torino, Italy}
\affiliation{DISAT, Politecnico di Torino, Corso Duca degli Abruzzi 24,
10129 Torino, Italy}

\author{Elena Losero}
\author{Ivano Ruo Berchera}
\affiliation{Quantum metrology and nano technologies division,  INRiM,  Strada delle Cacce 91, 10153 Torino, Italy}

\author{Stefano Pirandola}
\affiliation{Department of Computer Science, University of York, York YO10 5GH, United Kingdom}

\author{Marco Genovese}
\affiliation{Quantum metrology and nano technologies division,  INRiM,  Strada delle Cacce 91, 10153 Torino, Italy}
\begin{abstract}
The final goal of quantum hypothesis testing is to achieve quantum advantage over all possible classical strategies. In the protocol of quantum reading this advantage is achieved for information retrieval from an optical memory, whose generic cell stores a bit of information in two possible lossy channels. For this protocol, we show, theoretically and experimentally, that quantum advantage is obtained by practical photon-counting measurements combined with a simple maximum-likelihood decision. In particular, we show that this receiver combined with an entangled two-mode squeezed vacuum source is able to outperform any strategy based on statistical mixtures of coherent states for the same mean number of input photons. Our experimental findings demonstrate that quantum entanglement and simple optics are able to enhance the readout of digital data, paving the way to real applications of quantum reading and with potential applications for any other model that is based on the binary discrimination of bosonic loss.
\end{abstract}
\maketitle

\textbf{Introduction.}--~In the vast panorama of quantum technologies~\cite{Nielsen00,Bouwmeester00}, the most practical area is arguably that of quantum sensing, well developed with both discrete~\cite{Degen17} and continuous variable systems~\cite{SensingReview,RuoBerchera19,Genovese16}. In this area, quantum metrology~\cite{Braunstein94} deals with the estimation of unknown parameters encoded in a state or a physical transformation, while quantum hypothesis testing~\cite{Helstrom76} deals with the discrimination of a discrete set of states~\cite{Chefles98,Chefles00,Barnett09,Bergou10} or quantum channels~\cite{Lloyd08,Tan08}. In particular, the problem of quantum channel discrimination~\cite{SensingReview,Kitaev97} is known to have a very rich theoretical structure due to its inherent double optimization nature, which involves finding both the best input states and the optimal output measurements.

In 2011, Ref.~\cite{qReading} modeled the information retrieval from an optical memory as a problem of bosonic channel discrimination. In fact, a memory cell can be represented as a reflector (e.g., a beam splitter) with two possible values of the reflectivity, which is equivalent to considering two possible lossy channels acting on the incoming photons. In this scenario, one can show that the use of a quantum source of light (and, in particular, entangled) can sensibly boost the retrieval of information from the cell with respect to classical input states, i.e., having positive-P representations~\cite{Prepres,Prepres2}.

The idea of quantum reading has been further explored in a series of papers (e.g., see Refs.~\cite{qread1,qread2,qread3,qread5b,qread6,qread7} among others). A preliminary experiment~\cite{qread4} was performed for a perfect fully-unitary variant of the protocol, where zero discrimination error was achieved by analyzing the coincidences at the two outputs of the beam-splitter cell. For such an ideal unitary discrimination no entanglement is needed. However, in a realistic scenario, only one output of the cell is available for detection, so that the process is clearly non-unitary and must be described by a lossy quantum channel (as in the original proposal). For this reason, a truly quantum reading experiment has yet to be performed.
\begin{figure}[b]
\vspace{-0.5cm}
\includegraphics[width=0.35\textwidth]{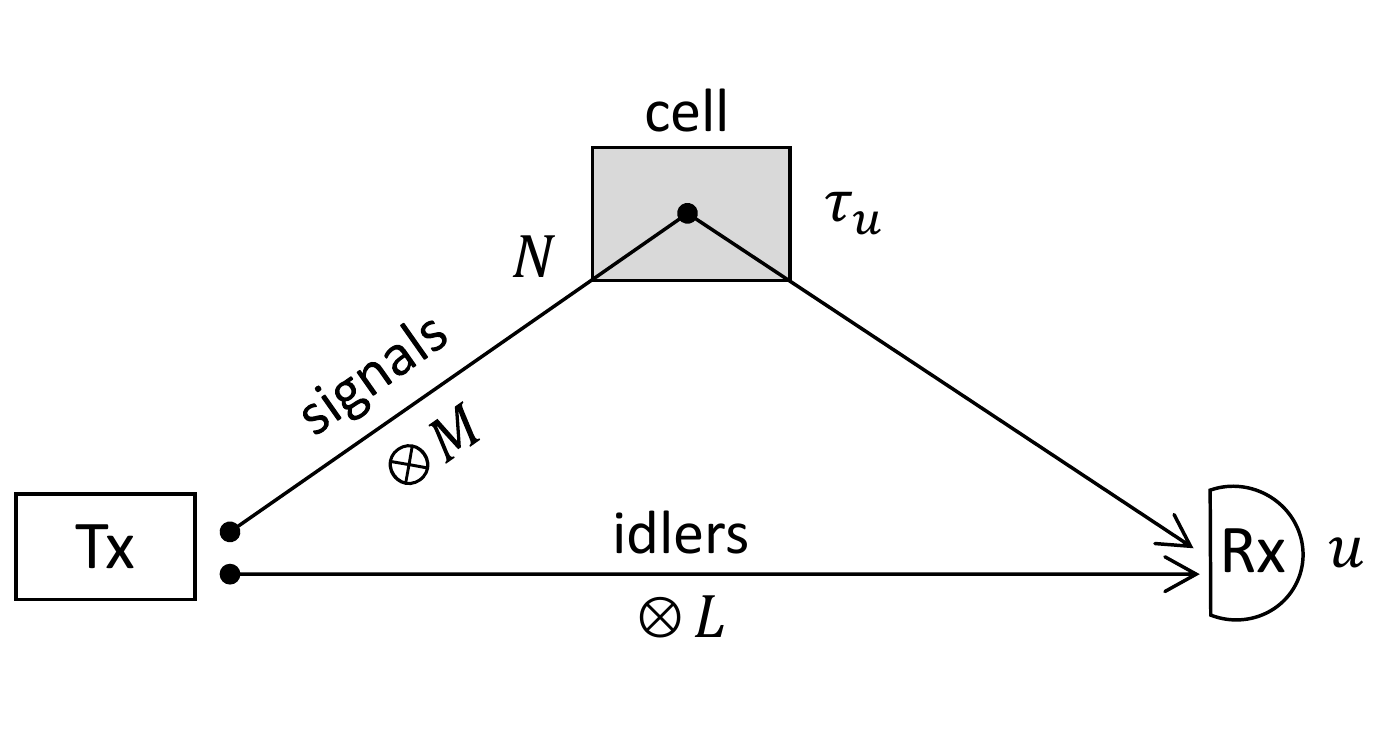}
\vspace{-0.3cm}
\caption{\emph{Quantum reading of a memory cell}. A memory cell encodes a bit $u$ in a lossy channel with transmissivity $\tau_u$. The cell is read by a transmitter (Tx) which irradiates $M$ signal modes and $N$ mean total photons over the cell, plus extra $L$ idler modes sent to the output. The receiver (Rx) performs a generally-joint measurement of signals and idlers, decoding the bit $u$ up to some error probability $p_{err}$. Quantum reading corresponds to using a quantum source of light for the Tx, so that we outperform any classical source in the readout of the bit. The scheme can be realized in reflection or in transmission, as done in our experiment.} \label{QRfig}
\end{figure}

In this work, we experimentally demonstrate the original protocol of quantum reading~\cite{qReading} showing that a two-mode squeezed vacuum state (TMSV) ~\cite{Weedbrook12,Braunstein05} is able to outperform any classical state in retrieving information from an absorbing layer in a coated glass-slide, mimicking the memory cell. Remarkably, this advantage is achieved without resorting to any complicated Helstrom-like measurement~\cite{Helstrom76,QCB,QCBGauss}, but just resorting to photon counting of the output followed by a maximum likelihood decision. Quantum advantage is proven notwithstanding the presence of more than 20\% experimental loss. This robustness to losses and the simplicity of detection scheme pave the way to possible real applications of quantum-reading in a next future.

\smallskip
\textbf{Theoretical model}.--~Let us store a bit $u=\{0,1\}$ in a memory cell by means of two equiprobable lossy channels, $\mathcal{E}_{0}$ and $\mathcal{E}_{1}$,  with transmissivities $\tau_0$ and $\tau_1$. Recall that a lossy channel with transmissivity $\tau$ corresponds to the following input-output transformation of the field operator $\hat{a} \rightarrow \sqrt{\tau}\hat{a}+i\sqrt{(1-\tau)}\hat{v}$, where $\hat{v}$ describes an environmental vacuum mode \cite{Meda17}. To retrieve the bit, consider a transmitter and a receiver. The transmitter irradiates $M$ signal  modes over the cell, for a total of $N$ mean photons, and also sends additional $L$ idler modes directly to the output. The receiver measures the transmitted signals and the idlers, guessing the classical bit $u$ up to an error probability $p_{err}$ (see Fig.~\ref{QRfig}).

Assuming an optimal measurement at the receiver, the minimization of $p_{err}$ over all transmitters with fixed signal energy $N$ is difficult to solve. If we restrict the analysis to classical transmitters, described by a state with positive P-representation (mixture of coherent states), then the minimum error probability is given by~\cite{qReading}
\begin{equation}
p^{cla}_{err}\geq\mathcal{C}(N,\tau_0,\tau_1):=\frac{1-\sqrt{1-e^{-N(\sqrt{\tau_1}-\sqrt{\tau_0})^2}}}{2}. \label{cl}
\end{equation} 
Equivalently, the maximum information accessible to classical transmitters cannot exceed the bound $1-H(\mathcal{C})$, where $H(\cdot)$ denotes the binary Shannon entropy~\cite{CoverThomas}.

Consider now a multi-mode quantum transmitter in a tensor product of $M$ TMSV states $|\mathrm{TMSV}\rangle_{S,I}^{\otimes M}$. Each TMSV state irradiates $\bar{n}$ mean photons per mode and describes an entangled pair of signal (S) and idler (I) modes, so that we have a total of $M$ signals and corresponding $L=M$ idlers. Let us assume that $\bar{n}$ is chosen such that $M\bar{n}=N$ mean photons are globally irradiated over the cell. Then, for sufficiently large $M$, it is possible to show that the error probability $p_{err}$ goes below the classical bound $\mathcal{C}$. In terms of the gain
\begin{equation}\label{eq:gain}
G=1-H(p_{err})-[1-H(\mathcal{C})],
\end{equation}
one can show that $G$ may approach $1$, meaning that the quantum transmitter retrieves all the information while the bit cannot be read by any classical strategy~\cite{qReading}.

In the following we show that a similar result can be achieved performing a photon counting measurement at the output and a maximum likelihood decision, in the place of the unspecified optimal receiver. Note that quantum advantage has been demonstrated by photo counting measurement strategies for parameter estimation~\cite{Tapster1991,Brida2010,Samantaray2017,Moreau2017,sabines19,Ortolano19,Losero18}, where the goal is to estimate the value of a continuous parameter $\tau$.  In that case, in fact, it can be proven that suitable quantum resources and photon counting measurements allow one to reach the ultimate (non-adaptive) quantum limits in precision~\cite{paris2007, adesso2009, Nair18, Losero18}. However, for the discrete-case considered here, i.e., for a problem of binary channel discrimination, such a proof has not been given and the effective performance of photon counting has not been investigated yet. 

\textbf{Photon counting strategy.--~}When photon counting measurements are performed over the signal and idler modes of a bipartite state $\rho$, the output is a classical random variable $\textbf{n}=(n_S,n_I)$, distributed as $p(\textbf{n})=\langle n_S,n_I|\rho |n_S,n_I\rangle$, where $| n_k\rangle$ is the eigenstate with eigenvalue $n_k$ of the number operator $\hat{n}_k=\hat{a}_k^{\dagger} \hat{a}_k$ of the field and $k=S,I$. 
The effect of a lossy channel $\mathcal{E}_\tau$ on the signal mode of a bipartite state is to combine its initial photon distribution $p_0(\textbf{n})$ with a binomial distribution $\mathcal{B}(n_S'|n_S,\tau)$ with $n_S$ trials and success probability $\tau$, so that the outcome $\textbf{n}$ will be distributed according to
\begin{equation}
   p(\textbf{n}|\tau)=\sum_{m=n_S}^\infty p_0(m,n_I)\mathcal{B}(n_S|m,\tau).
\end{equation}

Let us suppose that $\textbf{n}$ is the outcome of photon-counting measurements after a lossy channel with unknown transmissivity $\tau_u$ (for $u=0,1$). Using Bayes' theorem, the conditional probability of $\tau_u$ is given by
\begin{equation}
p(\tau_u|\textbf{n})=\frac{p(\textbf{n}|\tau_u)p(\tau_u)}{p(\textbf{n})}=\frac{p(\textbf{n}|\tau_u)}{p(\textbf{n}|\tau_0)+p(\textbf{n}|\tau_1)}\label{pb},
\end{equation}
where the last equality follows from the condition of equi-probable channels, $p(\tau_u)=1/2$. To assign a value to the recovered bit, the optimal strategy is to choose the value $u=0,1$ such that
$u=\arg\max_{u}p(\tau_{u}|\mathbf{n})$. Because $p(\tau_u)$ is uniform, this is equivalent to a maximum likelihood decision, i.e., to choose $u=\arg\max_{u}p(\mathbf{n}|\tau_{u})$.

The corresponding error probability will be given by $p_{err}(\tau_0,\tau_1|\mathbf{n})=\min_{u}p(\tau_u|\mathbf{n})$. Therefore, by averaging over the distribution of the outcomes $p(\textbf{n})$, we may write the following expression for the mean error probability
\begin{align}
p_{err}(\tau_0,\tau_1) &= \sum_{\mathbf{n}}\min_{u}p(\tau_u|\mathbf{n}) p(\mathbf{n}) \nonumber \\ &=\frac{1}{2}\sum_{\mathbf{n}}\min_{u}p(\mathbf{n}|\tau_u). \label{ps}
\end{align}

The error probability above describes the performance achievable by a photon-counting receiver in the reading scenario of Fig.~\ref{QRfig} where the transmitter irradiates a generic bipartite state. In general, the formula can be applied to a transmitter with arbitrary $M$ and $L$ by considering an $M+L$ vectorial variable $\mathbf{n}$. Let us now apply this analysis to evaluate the corresponding performances with classical and quantum states. Without loss of generality, in the following we assume that $\tau_0 <\tau_1$.

The photon counting performance with a classical transmitter, i.e., described by a state with positive P-representation, is optimized by the use of a single signal mode ($M=1$ and $L=0$) with $N$ mean photons, whose photon number statistics is a Poisson distribution $\mathcal{P}_{N}(n)$. It is easy to show that there is a threshold value $n^{th}:=N (\tau_1-\tau_0)/\log(\tau_1/\tau_0)$ such that, for every $n \leq n^{th}$, one has $\mathcal{P}_{N}(\tau_0|n)>\mathcal{P}_{N}(\tau_1|n)$, thus the value $\tau_0$ is chosen. The error probability will be given by
\begin{equation}
p_{err}^{cla,phc}(\tau_{0},\tau_{1})=\frac{1}{2}\left[  1-\frac{\gamma(\tau
_{0})-\gamma(\tau_{1})}{\lfloor n^{th}\rfloor!}\right],  \label{pl}%
\end{equation}
where $\lfloor x \rfloor$ is the floor of $x$, $\gamma(\tau_{u}):=\Gamma(\lfloor n^{th}+1\rfloor,N\tau_{u})$, and $\Gamma(x,y)$ is the incomplete gamma function. In other words, Eq.~(\ref{pl}) establishes a lower bound on the error probability that can be achieved by using classical transmitters and photon counting.

 \begin{figure*}
\includegraphics[ width=0.8\textwidth]{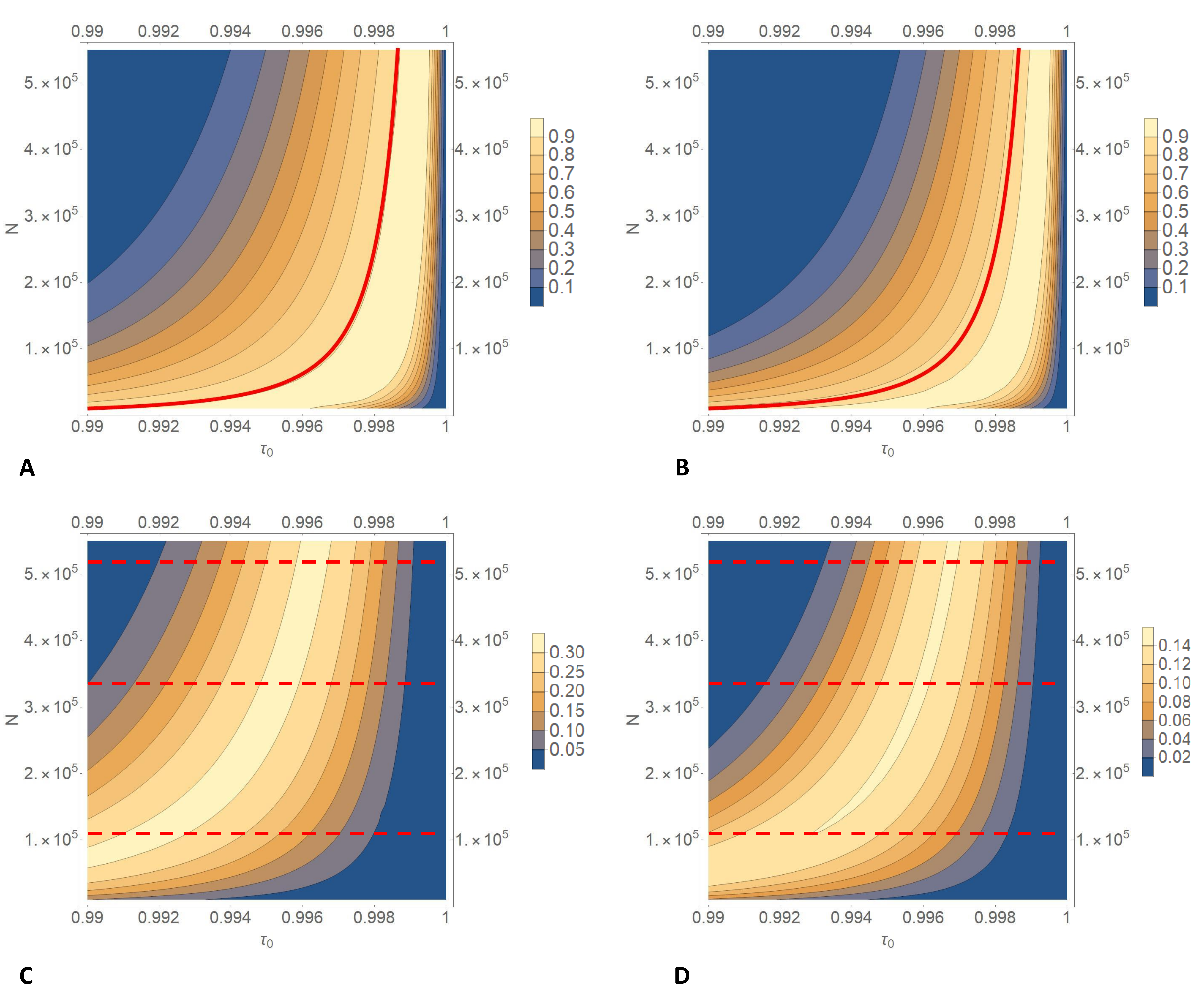}
\caption{\label{fig:cp} Information gain $G$ of quantum reading as a function of the lower transmissivity $\tau_0$ and total mean number of photons $N$ (higher transmissivity is set to $\tau_1=1$). The information gain is computed assuming a TMSV-state transmitter with large number of copies ($M \simeq 10^{13}$) and a receiver based on photon counting. In panel~\textbf{A}, the classical benchmark is the photon-counting performance with classical states of Eq.~(\ref{pl}). In panel~\textbf{B}, the benchmark is the optimal classical limit in Eq.~(\ref{cl}). In both panels, the red curve represent the MED strategy described in the text, marking the limit after which the channels are classically indistinguishable. In panels~\textbf{C} and~\textbf{D}, we consider the case of imperfect quantum efficiency $\eta=0.76$ for both the signal and idler systems (so that $\tau_u \rightarrow \eta \tau_u$ for $u=0,1$). We show the gain over the photon counting classical bound in panel~\textbf{C}, and the gain over the optimal classical limit in panel~\textbf{D}. In these panels, the dashed lines indicate the regions where experimental data were collected. These data points are those reported in Fig.~\ref{fig:exp_gain}.
}
\end{figure*}

Let us now study the photon-counting performance that is achievable by a quantum transmitter based on copies of TMSV states. We consider the transmitter's state $\vert \mathrm{TMSV}\rangle_{S,I}^{\otimes^{M}}$, where each signal-idler TMSV state $\vert \mathrm{TMSV}\rangle_{S,I}\propto\sum_{n} \sqrt{P_{\bar{n}}(n)}\vert n\rangle_{S}\vert n\rangle_{I}$ is maximally correlated in the number of photons and locally characterized by a single-mode thermal distribution $P_{\bar{n}}(n)=\bar{n}^n/(\bar{n} + 1)^{n+1}$. The product state $\vert \mathrm{TMSV}\rangle_{S,I}^{\otimes^{M}}$  preserves the perfect correlation between the total photon numbers, redefined as $\sum_{m=1}^{M} n^{(m)}_{S/I}\rightarrow n_{S/I}$, while the marginal distribution becomes multi-thermal $P_{N,M}(n_{S/I})$, with mean photon number N. Fixing $N$ and increasing $M$, this distribution becomes narrower and tends to a Poisson distribution $\mathcal{P}_N(n_{S/I})$ with mean occupation number $N/M\to 0$.

The presence of a memory cell with transmissivity $\tau_u$ on the signal path transforms the input joint probability $P_{N,M}(n_S,n_I)$ into the output probability distribution $P_{N,M}(n_S,n_I|\tau_u)= P_{N,M}(n_I) \mathcal{B}(n_S|n_I,\tau_u)$. Photon counting is then performed on both the signal and idler modes, and a maximum likelihood decision is finally taken. In fact, we can identify a threshold value
\begin{equation}
n_S^{th}=\left\{\frac{\log(\tau_1/\tau_0)}{\log[(1-\tau_0)/(1-\tau_1)]}+1\right\}^{-1}n_I,
\end{equation}
and choose $\tau_0$ if $n_S<n_S^{th}$, corresponding to the condition $P_{N,M}(n_S,n_I|\tau_0) > P_{N,M}(n_S,n_I|\tau_1)$. Otherwise we choose $\tau_1$. This strategy provides an error probability $p_{err}^{qua,phc}$ for the TMSV-based transmitter and the photon-counting receiver. We explicitly evaluate the performance of this strategy in the numerical study below.

\smallskip

\textbf{Theoretical predictions.--~}Numerical investigation shows a quantum advantage even with a single TMSV state. However, the described narrowing of the marginal distributions, resulting from the spread of the energy over an high number of copies $M$, makes the discrimination more effective, so this is the regime that we will consider and exploit in our experiment. 
 We have studied the following information gain $G=1-H(p_{err}^{qua,phc})-[1-H(p_{err}^{cla})]$, where we have assumed, for $p_{err}^{cla}$, either the optimal classical bound in Eq.~(\ref{cl}) or the classical photon counting bound of Eq.~(\ref{pl}).

As we can see from Fig.~\ref{fig:cp}, there is an evident information gain, which may approach the maximum value of $1$, meaning that, in certain regions the use of quantum resources allows the full recovery of the stored information, whereas no information could be retrieved by classical means.

In Fig.~\ref{fig:cp}(A,B) we see that, increasing the mean photon number, the maximum of the advantage shifts towards higher reflectivity $\tau_0$. Intuitively, this is explained by the fact that the gain becomes larger when classical strategies start to fail. For example, although non-optimal, another classical discrimination strategy can be to measure the mean photon number, that is either $N$ or $\tau_0N$ (assuming $\tau_1=1$). This approach of mean-energy-discrimination (MED) fails when the difference in the average photon counts becomes smaller than the noise associated with the Poisson fluctuations, i.e., when $\tau_0>1-N^{-1}$. The saturation of this inequality defines the red line in Fig.~\ref{fig:cp}. 

In fact, in Fig.~\ref{fig:cp}(A), this curve follows the contour lines of the plot, denoting the start of the maximum gain region. Of course, when $\tau_0$ is approaching $\tau_1=1$ there is no way to distinguish among the channels, neither classical nor quantum, and the information gain drops to zero. The competition between these two tendencies determines the maximum of the gain. When comparing with the optimal classical bound in Fig.~\ref{fig:cp}(B), the regions are in general narrower, and the maximum deviates from the MED curve. However, note that Eq.~(\ref{cl}) represents a theoretical lower bound which may be non-tight. 

The biggest limitation in an experimental realization of this procedure is given by photon losses of different nature, interaction with the environment and optical components, as well as the intrinsic quantum efficiency of the detectors. Their combined effect can be accounted with a unique coefficient, the detection efficiency $0 \leq \eta \leq 1$, that can be estimated with high precision in the characterization of the setup. This quantity expresses the fraction of generated photons that are actually detected. Moreover, in case of bipartite correlations, it may include the efficiency in detecting correlated photons, which can be lower than the efficiency in detecting the photons in a single arm. Its effect is indistinguishable from the effect of any other attenuator, such as the memory storing the value of a bit in its coefficient.

The composition property of two binomial processes implies that two consecutive pure-loss channels, $\mathcal{E}_{\eta}$ and $\mathcal{E}_{\tau}$, commute and their total effect is given the composite pure-loss channel $\mathcal{E}_{\eta\tau}$.  Due to this indistinguishability, the classical limits, in this scenario can be computed performing the substitution $\tau_u \rightarrow \eta\tau_u$ in Eq.~(\ref{cl}) and Eq.~(\ref{pl}), resulting in a decreased accuracy for discrimination. An equivalent way to obtain these classical limits is to consider the signal energy reduction caused by $\eta$, yielding the same result. When quantum-correlated systems are considered, however, aside from the energy reduction, an additional effect induced by losses is the worsening of the correlations, therefore decreasing the advantage that can be obtained. This drop in the gain can be seen from Figs.~\ref{fig:cp}(C-D), where the scenario with an efficiency $\eta=0.76$ is reported. The maximum gain is reduced to $\simeq 1/3$ or $\simeq 1/6$, depending on the classical benchmark considered. Still, this is a macroscopic amount of information due to the fact that it refers to gain \textit{per cell}.

\begin{figure}
\centering
  \includegraphics[trim={1cm 0.5cm 2cm 0.5cm}, clip, width=1\columnwidth]{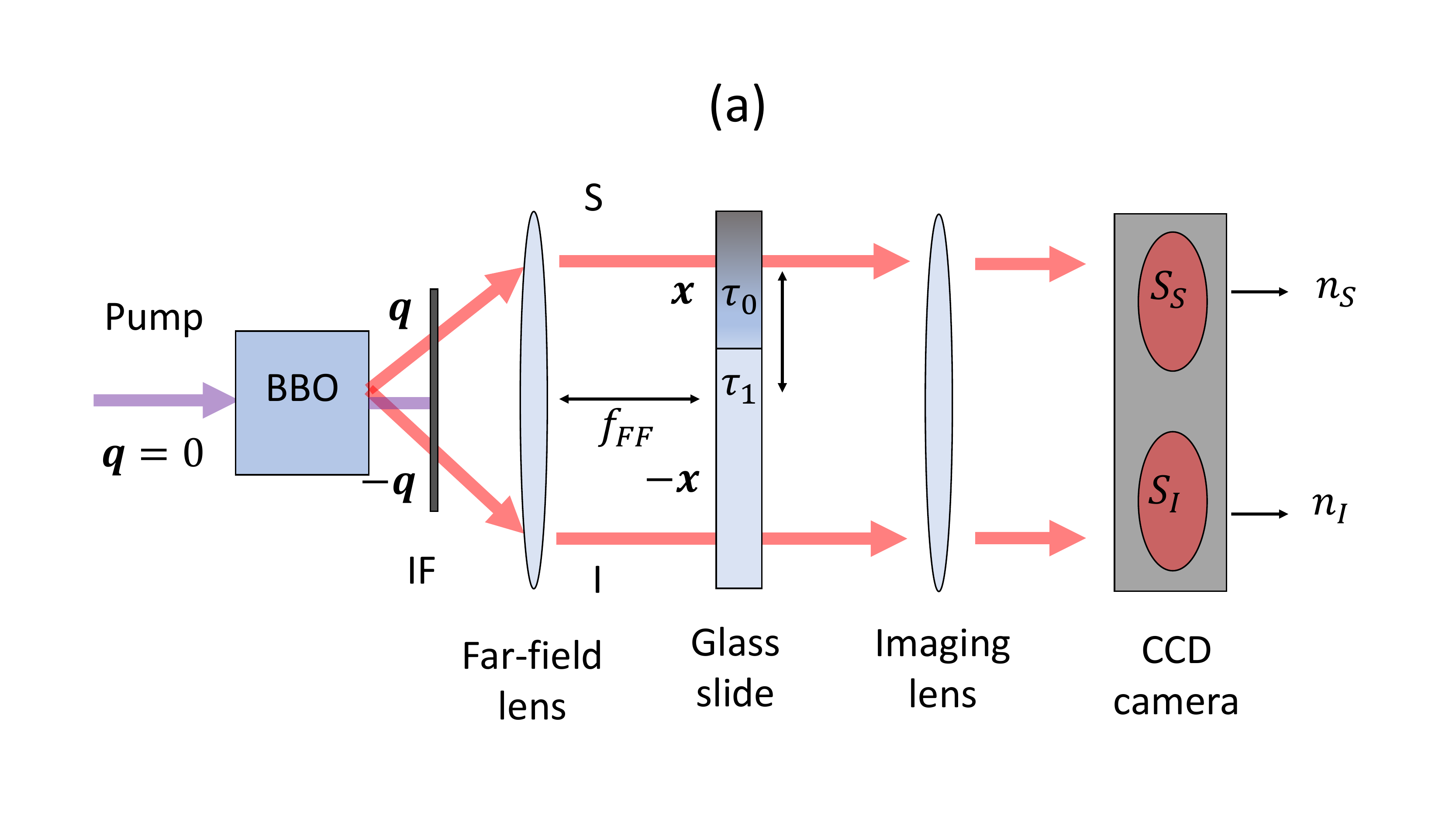}
  \includegraphics[trim={0cm 0.5cm 2cm 0.5cm}, clip, width=1\columnwidth]{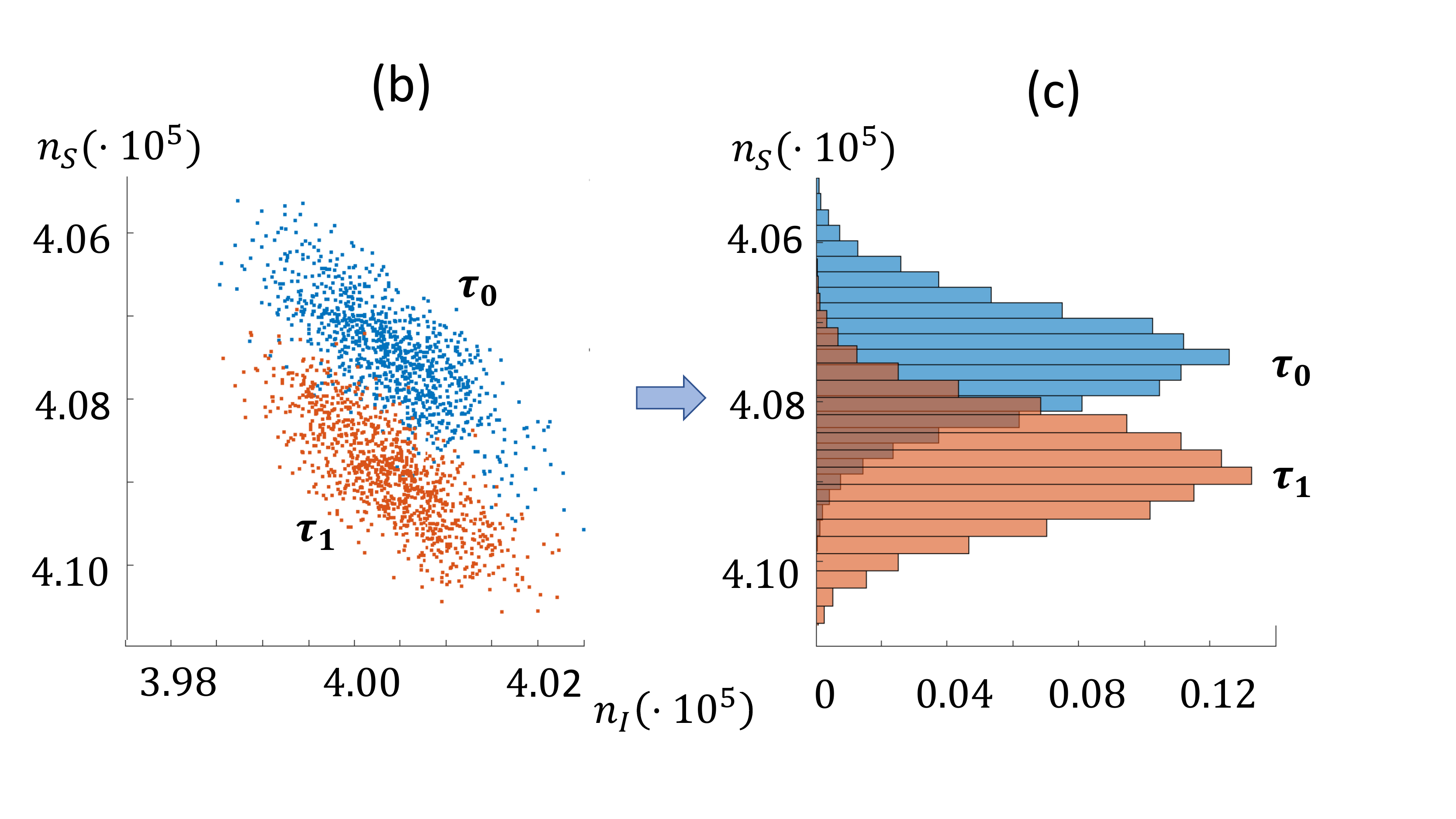}
\caption{
(a) Simplified schematic of the experimental set-up. In the BBO crystal the multi-mode TMSV source is generated. The signal beam passes through the memory cell investigated, whose transmissivity can be either $\tau_0$ or $\tau_1$ and is then detected in the $S_S$ region of the CCD camera. The idler beam goes directly to the $S_I$ region of the CCD. $n_S$ and $n_I$ are the total photon counts over $S_S$ and $S_I$. BBO: Type-II-Beta-Barium-Borate non linear crystal. IF: interferential filter ($800 \pm 20$nm). CCD: charge-coupled device camera. (b) $n_S$ in function of $n_I$, for 1000 frames. Blue dots correspond to $\tau_0 \sim 0.996$, while red dots corresponds to $\tau_1=1$. (c) $n_S$ relative frequency distribution for $\tau_0 \sim 0.996$ (blue histogram) and $\tau_1=1$ (red histogram).}
\label{fig:setup}
\end{figure}
\begin{figure}
 \includegraphics[trim={0cm 0cm 0cm 0cm}, clip, width=1\columnwidth]{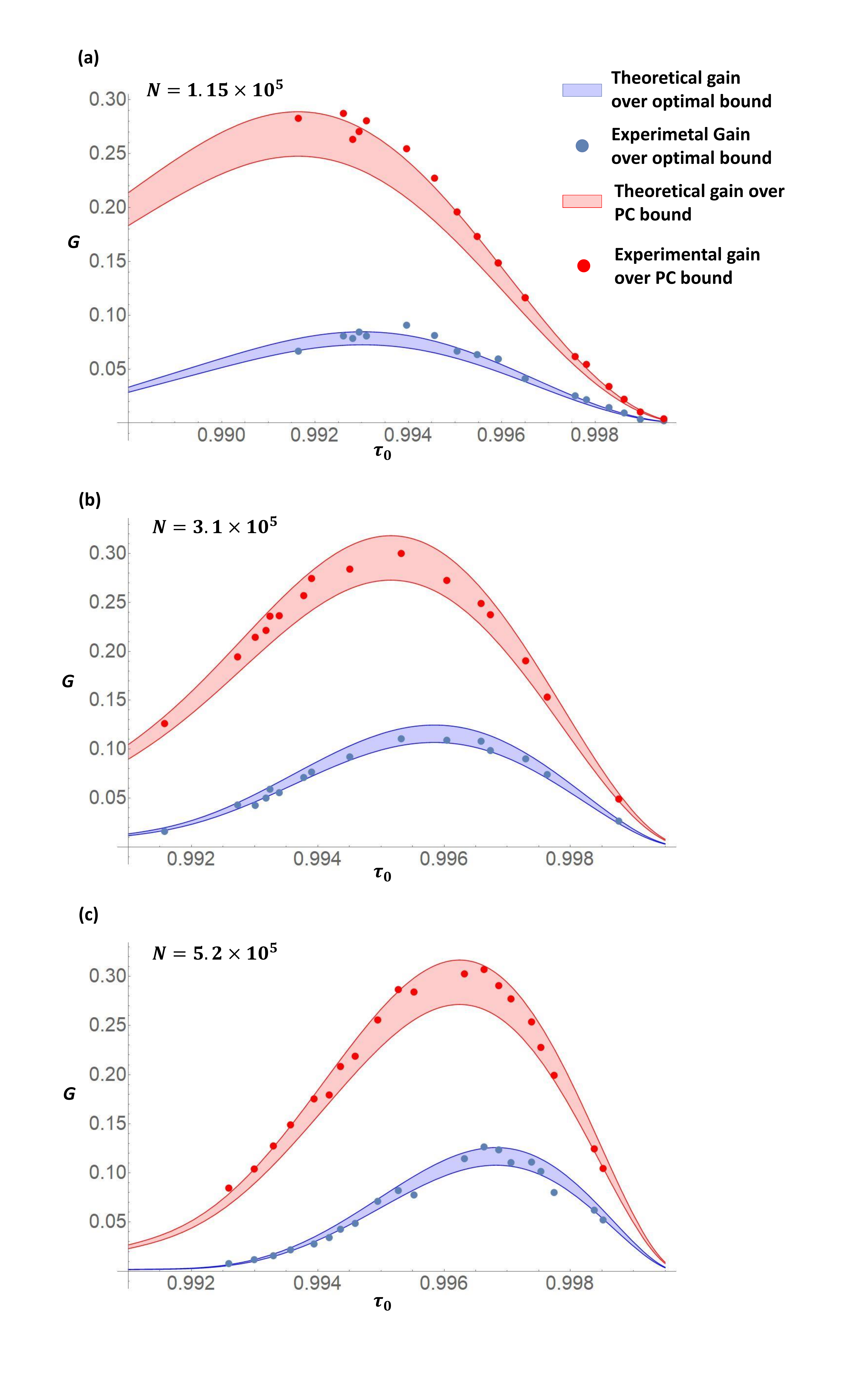}
\caption{Experimental gain $G$ of quantum reading (bits) as a function of the lower transmissivity $\tau_0$. The three panels refer to different mean photon number in the signal beam: (a) $N=1.15 \cdot 10^5$, (b) $3.1 \cdot 10^{5}$, and (c) $5.2 \cdot 10^{5}$. Blue data refers to the gain with respect to the classical optimal bound in Eq.~(\ref{cl}). Red data refers to the gain with respect to the classical photon-counting bound given in Eq.~(\ref{pl}), obtained from the marginal distribution of the signal. The experimental parameters, estimated independently in a calibration step, are the mean signal energy $N$, the detection efficiency of signal and idler channel $\eta_S$ and $\eta_I$ and the electronic noise $\nu_e$. Apart the value of $N$, which is intentionally different in the three panels, the other parameters are kept fixed to: $\eta_S=0.78$, $\eta_I=0.77$, and $\nu_e\sim10^4$.}
\label{fig:exp_gain}
\end{figure}

\smallskip
\textbf{Experimental results.--~}
 A scheme of the experimental set-up is reported in Fig.~\ref{fig:setup}(a). The multi-mode state $\vert \mathrm{TMSV}\rangle_{S,I}^{\otimes^{M}}$ is experimentally produced exploiting the spontaneous parametric down conversion process in a non linear crystal. We pump a $(1\mathrm{cm})^3$ type-II-Beta-Barium-Borate (BBO) crystal with a CW laser of $\lambda_p=405$nm and power of $100$mW. 
An interferential filter (IF) at $(800 \pm 20)$nm performs a spectral selection of the down-converted photons around the degenerate frequency ($\lambda_d=2 \lambda_p=810$nm). The correlation in momentum of two down-converted photons is mapped into spatial correlations at the back focal plane of a lens with $f_{FF}=1$cm focal length. This plane is then imaged to the detection plane by a second lens. 

The detector is a charge-coupled-device (CCD) camera (Princeton Instrument Pixis 400BR Excelon), working in linear mode, with high quantum efficiency (nominally $> 95\%$ at $810$nm) and few $e^{-}/(\mathrm{Pixel} \cdot \mathrm{Frame})$ of electronic noise. The physical pixels of the camera measure $13 \mu$m. A $12 \times 12$ hardware binning is performed on them, in order to lower the acquisition time and increase the read-out signal-to-noise ratio. The total photon counts $n_S$ and $n_I$ are obtained integrating the signal over the two spatially correlated detection areas $S_S$ and $S_I$, for signal and idler respectively. The total number of spatial modes collected is $M_s\sim 10^3$ and the temporal modes can be estimated to be $M_{t}  \sim  10^{10}$ (for a deeper discussion on these estimates see \cite{acoh1}). Since $ N_I \sim 10^5$, the mean occupation number is $N_I/(M_s \cdot M_t) \sim   10^{-8} \ll 1$, meaning that the marginal distributions are well approximated by Poissonian ones.

The memory cell is implemented inserting in the focal plane of the first lens a coated glass-slide with a deposition of variable transmission $0.990 < \tau_0 < 1$. The bit of information is stored in the presence ($\tau=\tau_0$) or absence ($\tau=\tau_1=1$) of the deposition. 

The effect at the base of the quantum enhancement can be visualized comparing Fig.~\ref{fig:setup}(b) and Fig.~\ref{fig:setup}(c). The joint distributions of $n_S$ and $n_I$ for $\tau_0$ and $\tau_1=1$, due to their squeezed shape, are less overlapped with respect to the marginal distributions of $n_S$ only, increasing their distinguishability.  Note that the squeezed shape the joint distributions of Fig.~\ref{fig:setup}(b) is purely due to quantum correlations and cannot be achieved by any classical source. \\

The parameters necessary for the subsequent analysis ($N$, $\tau_0$, $\eta_1$, $\eta_2$, electronic noise $\nu_{e}$) are estimated in a calibration phase. In particular, the channels efficiencies are estimated using the absolute calibration method presented in Refs.~\cite{etacalib,etacalib2,etacalib3}.
The error probability in the discrimination between $\tau_0$ and $\tau_1$ is evaluated on two sets of frames (10000 frames per set are acquired), one for each known value of the transmittance. 
For each frame we compute $P_{N,M}(n_S,n_I|\tau_u)$, using the values of the parameters estimated in the calibration, and we assign to the frame the value of $\tau_u$ that makes this probability higher. 
The comparison of the true  known  value  of $\tau_u$ over  each  set  with  the  guessed ones,  allows  estimating  the  error  frequency $p^{exp}_{err}$ for each set. 


The experimental gain $G$ evaluated from $p^{exp}_{err}$ is reported in Fig.~\ref{fig:exp_gain}, both with respect to the optimal classical bound (blue curves) and to the classical photon-counting bound (red curves). The three panels are obtained for a different number of photons in the signal beam, i.e., $N \sim 1.15 \cdot 10^{5}$, $3.1 \cdot 10^{5}$ and $5.2 \cdot 10^{5}$ respectively, corresponding to the sections lines in the theoretical Figs.~\ref{fig:cp}(C-D). In Fig.~\ref{fig:exp_gain}, the error bands on the theoretical curves have been obtained via numerical simulation. Experimental data show a good accordance with the theoretical model, with the majority of the data falling in the confidence region at $1$ standard deviation. 
In all three cases, we find a clear quantum advantage. In perfect accordance with theory, we find that the maximum gain increases with the mean signal energy but at the expenses of a narrowing of the region in which the quantum enhancement can be found.

\smallskip
\textbf{Conclusion.--~}
In this work we have provided an experimental demonstration of the quantum reading protocol, showing how entanglement is able to boost the retrieval of classical information from an optical memory cell, outperforming any classical strategy for the same number of input photons. We have shown, theoretically and experimentally, that quantum advantage can be achieved by means of a simple receiver strategy based on photon counting measurements followed by a maximum likelihood decision test. In this way, we were able to demonstrate values, for the quantum advantage, which are close to the performance originally foreseen by using optimal, but highly-theoretical, joint quantum measurements.

In our experiment, we considered the realistic scenario where only a single output from the cell is accessible for detection and we were able to show quantum advantage despite the presence of extra optical losses on both the signal and idler paths. Because of all these aspects, our results pave the way for a realistic and practical implementation of quantum reading techniques, whose implications go beyond the memory model and may involve spectroscopic applications. For instance, our results implicitly show the feasibility of a quantum-enhanced detection of absorbance at some frequency of a spectrum. Thus, this work represents a significant step in the progress of quantum technology, demonstrating the feasibility with easily accessible resources of a quantum scheme of huge practical interest. 

\smallskip
\textbf{Acknowledgments.--~}This work has been sponsored by the EU via ``Quantum readout techniques and technologies'' (QUARTET, Grant agreement No 862644). 


\end{document}